\documentclass[twocolumn,showpacs,nofootinbib]{revtex4-1}
\usepackage{epsfig,amssymb,amsmath,latexsym}
\usepackage{pifont}
\usepackage{hyperref}

\usepackage{graphicx}
\usepackage{dcolumn}
\usepackage{bm}
\usepackage{epsfig}
\usepackage{color}

\usepackage{color}
\usepackage{hyperref}

\hypersetup{
    colorlinks=true,
    linkcolor=red,
    citecolor=blue,
}

\def\fun#1#2{\lower3.6pt\vbox{\baselineskip0pt\lineskip.9pt
  \ialign{$\mathsurround=0pt#1\hfil##\hfil$\crcr#2\crcr\sim\crcr}}}
\def\simgt{\mathrel{\lower0.6ex\hbox{$\buildrel {\textstyle >}
 \over {\scriptstyle \sim}$}}}
\def\simlt{\mathrel{\lower0.6ex\hbox{$\buildrel {\textstyle <}
 \over {\scriptstyle \sim}$}}}

\def\bea{\begin{eqnarray}}
\def\eea{\end{eqnarray}}
\def\be{\begin{equation}}
\def\ee{\end{equation}}

\input epsf

\def\be{\begin{equation}}
\def\ee{\end{equation}}
\def\ba{\begin{eqnarray}}
\def\ea{\end{eqnarray}}

\newcommand{\no}{\noindent}

\begin{document}

\preprint{}

\title{Vainshtein mechanism after GW170817}

\author{Marco Crisostomi$^{1}$ \& Kazuya Koyama$^{2}$}
	
\affiliation{
$^1$Institut de physique th\'eorique, Universit\'e Paris Saclay CEA, CNRS, 91191 Gif-sur-Yvette, France \\
Astrophysics Department, IRFU, CEA, Universit\'e Paris-Saclay, F-91191, Gif-sur-Yvette, France \\
Laboratoire de Physique Th\'eorique, CNRS, Universit\'e Paris-Sud,
Universit\'e Paris-Saclay, 91405 Orsay, France \\
$^2$Institute of Cosmology \& Gravitation, University of Portsmouth, Dennis Sciama Building, Portsmouth, PO1 3FX, United Kingdom}

\date{\small \today}

\begin{abstract}
\no The almost simultaneous detection of gravitational waves and a short gamma-ray burst from a neutron star merger has put a tight constraint on the difference between the speed of gravity and light. In the four-dimensional scalar-tensor theory with second order equations of motion, the Horndeski theory, this translates into a significant reduction of the viable parameter space of the theory. Recently, extensions of Horndeski theory, which are free from Ostrogradsky ghosts despite the presence of higher order derivatives in the equations of motion, have been identified and classified exploiting the degeneracy criterium.
In these new theories, the fifth force mediated by the scalar field must be suppressed in order to evade the stringent Solar System constraints.
We study the Vainshtein mechanism in the most general degenerate higher order scalar-tensor theory in which light and gravity propagate at the same speed.  We find that the Vainshtein mechanism generally works outside a matter source but it is broken inside matter, similarly to beyond Horndeski theories. This leaves interesting possibilities to test these theories that are compatible with gravitational wave observations using astrophysical objects.   
\end{abstract}


\maketitle
\noindent
{\bf Introduction.}
Einstein's theory of General Relativity (GR) has been proven successful over many years of experimental tests, ranging from sub-millimetre scale tests in the laboratory to Solar System tests and consistency with gravitational wave emission by binary pulsars and black holes. However, the standard model of cosmology is based on a huge extrapolation of our limited knowledge of gravity as GR has not been tested independently on galactic and cosmological scales. The discovery of the late time acceleration of the Universe provided therefore motivations to test gravity on cosmological scales and investigate modified theories of gravity (see reviews \cite{Clifton:2011jh, Joyce:2014kja, Koyama:2015vza}).  

In this respect, scalar-tensor theories of gravity play a special role since they represent the simplest modification in terms of additional degrees of freedom (dof) compared to GR, i.e. a single one.
In four-dimensional spacetime, the most general scalar-tensor theory with second order equations of motion (eom) was derived by Horndeski in 1974 \cite{Horndeski:1974wa}, and later rediscovered in the context of the so-called (covariant) galileon theories \cite{Nicolis:2008in, Deffayet0, Deffayet1, Deffayet2, Kobayashi:2011nu}. 
The requirement of second order eom was pursued to easily avoid the occurrence of Ostrogradsky instabilities \cite{Woodard:2015zca}, however, although sufficient, this condition is not necessary \cite{Motohashi:2016ftl, Klein:2016aiq, Crisostomi:2017aim}. 
In the recent years, there have been several attempts to construct healthy theories that relax this condition, exploiting transformations of the metric \cite{Zumalacarregui:2013pma, Gleyzes:2014dya}. This brought initially to the class of beyond Horndeski theories \cite{Gleyzes:2014dya, Gleyzes:2014qga}. Some of the Lagrangians in this class were shown to be related to the Horndeski ones by a disformal transformation \cite{Gleyzes:2014dya, Gleyzes:2014qga}, thus they are manifestly free from the Ostrogradsky ghost. However, it was not clear whether arbitrary combinations of them were still free from the ghost\footnote{Ref.~\cite{Deffayet:2015qwa} showed that the eom of beyond Horndeski theories could be rewritten into a system of equations that contain at most second-order time derivatives. However, this fact alone does not guarantee the propagation of 3 dofs \cite{Langlois:2015cwa}.}. A breakthrough in the subject came with the works of Refs.~\cite{Langlois:2015cwa, Langlois:2015skt}, which developed a general method to identify the degeneracy conditions that remove the Ostrogradsky ghost, despite the appearance of higher derivatives in the eom. Based on the degeneracy criterium, the viable sub-classes of beyond Horndeski theories were identified in \cite{Langlois:2015cwa, Crisostomi:2016tcp}. In addition, a larger class of new degenerate higher order scalar-tensor theories propagating up to 3 dofs was identified and classified up to cubic order in the second order derivative of the scalar field \cite{Langlois:2015cwa, Crisostomi:2016czh, Achour:2016rkg, BenAchour:2016fzp}. At the present time, the theories in \cite{BenAchour:2016fzp} represent the most general degenerate scalar-tensor theories propagating 3 dofs\footnote{Recently, new chiral scalar-tensor theories which break parity in the gravity sector were introduced \cite{Crisostomi:2017ugk}, however they propagate more than 3 dof away from the unitary gauge.}.

In August 2017, the first detection of a neutron star merger was made by LIGO and VIRGO detectors \cite{TheLIGOScientific:2017qsa}. The most striking discovery was the detection of the electromagnetic counter part SSS17a \cite{Coulter:2017wya, GBM:2017lvd,Murguia-Berthier:2017kkn}, which marked the start of multi-messenger astronomy. This event also gave a huge impact on the community studying modifications of gravity. The almost simultaneous detection of a short gamma-ray burst put extremely tight constraints on the difference between the speed of light and gravity. The constraint is given by $c_{GW}^2/c^2 -1 < 10^{-15}$, where $c_{GW}$ is the speed of gravitational waves and $c$ is the speed of light \cite{Monitor:2017mdv}. This has a significant implication for scalar-tensor theories as discussed in \cite{Lombriser:2015sxa, Lombriser:2016yzn, Bettoni:2016mij,Creminelli:2017sry, Sakstein:2017xjx, Ezquiaga:2017ekz, Baker:2017hug, Arai:2017hxj}. 

The scalar degrees of freedom mediates a fifth force, which is strongly constrained by precision tests of gravity at Solar System scales. Any dark energy and modified gravity model involving scalar fields should accommodate a mechanism to suppress the scalar interaction with visible matter on small scales, in order for them to be relevant only on cosmological scales. One of the oldest possibility is the Vainshtein mechanism \cite{Vainshtein}, originally discovered in the context of massive gravity (see \cite{Babichev:2013usa} for a review). This mechanism naturally appears also in Horndeski theories \cite{Kimura:2011dc, Koyama:2013paa} due to the presence of non-linear derivative interactions. In this paper, we study the consequence of the constraint on the difference between the speed of light and gravity on the Vainshtein mechanism, for the much wider class of theories \cite{BenAchour:2016fzp}, and assess whether the Vainshtein mechanism is still operating in these models or not.

\noindent
{\bf Scalar tensor theory.} To understand the implication of the gravitational wave constraint on the theories studied in  \cite{BenAchour:2016fzp}, it is sufficient to look at the effective action for linear cosmological perturbations derived in \cite{Langlois:2017mxy}. There, the deviation of $c_{GW}$ from $c$ was parametrised in term of a quantity called $\alpha_T$, explicitly given in terms of the free functions present in the theories studied in \cite{BenAchour:2016fzp}. Requiring $\alpha_T=0$ completely excludes the theories cubic in second derivatives of the scalar field, and constrains the quadratic theories in the following way\footnote{We thank David Langlois and Karim Noui for discussions on this point.}.
The most general Lagrangian quadratic in second derivatives of the scalar field reads
\cite{Langlois:2015cwa, Crisostomi:2016czh, Achour:2016rkg}
\be
{\cal L}_{\text{tot}}\,=\,\sum_{i=1}^5{\cal L}_i+{\cal L}_R \,, \label{ESTlag}
\ee
where
\begin{align}
{\cal L}_1 [A_1] &= A_1(\phi,\,X) \phi_{\mu \nu} \phi^{\mu \nu} \,,
\label{A1}
\\
{\cal L}_2 [A_2] &= A_2(\phi,\,X) (\Box \phi)^2 \,, 
\label{A2}
\\
{\cal L}_3 [A_3] &= A_3(\phi,\,X) (\Box \phi) \phi^{\mu} \phi_{\mu \nu} \phi^{\nu} \,, 
\label{A3}
\\
{\cal L}_4 [A_4] &= A_4(\phi,\,X)  \phi^{\mu} \phi_{\mu \rho} \phi^{\rho \nu} \phi_{\nu} \,, 
\label{A4}
\\
{\cal L}_5 [A_5] &= A_5(\phi,\,X)  (\phi^{\mu} \phi_{\mu \nu} \phi^{\nu})^2\,, 
\label{A5}
\end{align}
while
\begin{equation}
{\cal L}_R[G] = G(\phi,\,X) R \,,
\label{R}
\end{equation}
is a non-minimal coupling with gravity. We defined $\phi_{\mu} = \nabla_{\mu} \phi, \phi_{\mu \nu} = \nabla_{\mu} \nabla_{\nu} \phi$ and $X = \phi^{\mu} \phi_{\mu}$.  The functions $G$, $A_i$ are arbitrary functions of $\phi$ and  $X$ but for simplicity, and without any loss of generality, we will only consider them to be functions of $X$. For the Lagrangian (\ref{ESTlag}), the speed of gravitational waves computed from linear tensor perturbations around the cosmological background was firstly given in \cite{deRham:2016wji}:
\begin{equation}
c_{GW}^2 = \frac{G}{G -  X A_1}, \label{GWcond}
\end{equation}
where we now set $c=1$. Note that the Vainshtein mechanism is not able to screen
deviations in the speed of light and gravity \cite{Jimenez:2015bwa}. 

The Horndeski theory is given by the following choice of functions
\begin{equation}
A_1=-A_2= -2 G_{X}\,, \quad A_3=A_4=A_5=0,
\label{HOR}
\end{equation}
where $G_{X} = d G/dX$. Therefore, equation (\ref{GWcond}) implies that $G_{X}$ needs to be tuned to be small as $X G_{X}/G < 10^{-15}$ at least at the vicinity ($<40$ Mpc) of the Solar System today ($z<0.01$) \cite{Lombriser:2015sxa, Lombriser:2016yzn, Bettoni:2016mij,Creminelli:2017sry, Sakstein:2017xjx, Ezquiaga:2017ekz, Baker:2017hug, Arai:2017hxj}. It is still possible however to consider a highly tuned function so that $G_{X} =0$ only today meanwhile playing a role at $z>0$.

For the more general class of theories described by (\ref{ESTlag}), Eq.~(\ref{GWcond}) simply implies that the condition  
\be
A_1=0 \,,
\ee
needs to be satisfied to ensure $c_{GW}=1$ \cite{Ezquiaga:2017ekz} and this is what we will assume in this paper.   

To satisfy the degeneracy conditions that remove the Ostrogradsky ghost, the other functions should satisfy the following relations
\bea
\label{degcond}
A_2&=&0\,, \qquad A_5 = 
\frac{A_3}{2 G}(4 G_X + A_3 X)\,, \\
A_4&=& -\frac{1}{8 G}\left[
8 A_3 G - 48 G_{X}^2 - 8 A_3 G_{X} X + A_3^2 X^2
\right] \,, \nonumber
\eea
whereas $G$ and $A_3$ are left free.
This theory, with two free functions, is a subset of the class called N-I in \cite{Crisostomi:2016czh} and Ia in \cite{Achour:2016rkg} with $A_1=0$. Finally, we assume that matter is minimally coupled to the metric~$g_{\mu \nu}$. 

\noindent
{\bf Vainshtein mechanism.}
For the purpose of studying the Vainshtein mechanism in cosmology, we consider a cosmological background with a time dependent scalar field $\phi=\phi_0(t)$ and study the deviations around it, namely
\begin{equation}
ds^2 = -(1+ 2 \Phi(t, x^i)) dt^2 + a(t)^2 (1+ 2 \Psi(t, x^i)) 
\delta_{ij} dx^i dx^j,
\end{equation}
with $\phi = \phi_0(t) + \pi(t,x^i)$. The distinctive feature of the Vainshtein mechanism is that derivative self-interactions of the scalar field become large around a matter source and screen its effect. To identify the relevant terms describing the Vainshtein mechanism, we expand the equations of motion in terms of the fluctuations, using the following assumptions \cite{Kimura:2011dc, Koyama:2013paa}: the fields $\pi$, $\Phi$ and $\Psi$ are small, hence we neglect higher order interactions containing the metric perturbations $\Phi$ and $\Psi$, as well as terms containing higher order powers of the scalar field fluctuation $\pi$ and its first derivatives. On the other hand, we keep all terms with second or higher order spatial derivatives of perturbations, and will provide the necessary self-interactions to realize the Vainshtein mechanism. We will work with quasi-static approximations and ignore the time derivatives of the perturbations compared with the spatial derivatives. Note that we need to keep time derivatives for the terms containing second or higher order spatial derivatives in order to be consistent with the expansion scheme. With these assumptions, we obtain the following equations describing the dynamics of these perturbations:
\begin{align}
& {\cal G}_T \nabla^2 \Psi + {\cal G}_{T \Phi} \nabla^2 \Phi + a_2 \nabla^2 \pi
+a_{2}^t \nabla^2 \dot{\pi}  \\
& + b_2^{a} (\nabla^2 \pi)^2 + b_2^{b} (\nabla_{i} \nabla_j \pi)^2
+b_2^{c} (\nabla^i \pi)(\nabla_i \nabla^2 \pi) = a^2 \delta \rho, \nonumber
\end{align}
\begin{align}
&{\cal F}_T \nabla^2 \Psi - {\cal G}_T \nabla^2 \Phi + a_1^t \nabla^2 \dot{\pi} \nonumber\\
&+ b_1 (\nabla_i \nabla_j \pi)^2 
+b_1 (\nabla^i \pi)(\nabla_i \nabla^2 \pi) =0,
\end{align}
\begin{align}
&a_0 \nabla^2 \pi + a_0^t \nabla^2 \dot{\pi} + a_0^{tt} \nabla^2 \ddot{\pi} \nonumber\\
&+a_1 \nabla^2 \Psi + 2 a_1^t \nabla^2 \dot{\Psi} 
+a_3 \nabla^2 \Phi - 2 a_2^t \nabla^2 \dot{\Phi} \nonumber\\
&+ b_0^a (\nabla^2 \pi)^2 
+b_0^{b} (\nabla_i \nabla_j \pi)^2
+ b_0^{c} (\nabla^i \pi)(\nabla_i \nabla^2 \pi) \nonumber\\ 
& +b_0^t (\nabla^2 \dot{\pi} )(\nabla^2 \pi) 
+ 2 b_0^t (\nabla_i \nabla_j \dot{\pi}) (\nabla^i \nabla^j \pi) \nonumber\\
&+b_0^t (\nabla^i \dot{\pi}) (\nabla_i \nabla^2 \pi)
+2 b_0^t (\nabla^i \pi)(\nabla_i \nabla^2 \dot{\pi}) \nonumber\\
& +2 b_1 (\nabla^2 \Psi)(\nabla^2 \pi) + 2 b_1
(\nabla^i \nabla^2 \Psi) (\nabla_i \pi) + b_3 (\nabla^2 \Phi)(\nabla^2 \pi) \nonumber\\
& -4 b_2^a
(\nabla_i \nabla_j \Phi) (\nabla^i \nabla^j \pi) - b_2^c
(\nabla^i \nabla^2 \Phi) (\nabla_i \pi)  \nonumber\\
&+ c_0^a (\nabla^2 \pi)^3 + 2 c_0^a (\nabla_i \nabla_j \pi)^3 + c_0^{b} (\nabla_i \nabla_j \pi)^2 (\nabla^2 \pi)  \nonumber\\
&+ c_0^c (\nabla^i \pi)(\nabla_i \nabla_j \pi)(\nabla^j \nabla^2 \pi) + c_0^c (\nabla_i \pi) 
(\nabla^i \nabla^2 \pi) (\nabla^2 \pi)  \nonumber\\
&+ 2 c_0^c (\nabla^i \pi)(\nabla_i \nabla_k \nabla_j \pi)
(\nabla^k \nabla^j \pi)  \nonumber\\
&+ c_0^c (\nabla^i \pi)(\nabla^j \pi)(\nabla^2
\nabla_i \nabla_j \pi)
 =0, 
\end{align}
where $\delta \rho$ is the matter source and $\nabla_i$ is the spatial derivative with respect to $\delta_{ij}$. We do not give explicit expressions for these coefficients here as they are not important for our purpose. These equations contain up to the fourth order derivatives. This provides the extension of the non-linear operators identified in Horndeski theories around the cosmological background \cite{Kimura:2011dc}. Although the equations of motion contain higher order derivatives, it is still possible to reduce the system to the second order: we will demonstrate this explicitly for the spherically symmetric solutions.
   
\noindent
{\bf Spherically symmetric solutions.}
We now consider spherically symmetric solutions where the perturbations depend only on the radial coordinate and time. The three equations can be integrated once and we can solve $\Phi'$ and $\Psi'$ in terms of the scalar field perturbations $\pi'$, where the prime indicates the derivative with respect to $r$. The solutions for $\Phi$ and $\Psi$ have the following structure
\begin{align}
\Phi' & = \alpha^a \pi' + \alpha^b \dot{\pi}' + \alpha^c \pi'^2 + \alpha^d \pi' \pi'' + \beta^e M(t, r), \nonumber\\
\Psi' & = \beta^a \pi' + \beta^b \dot{\pi}' + \beta^c \pi'^2 + \beta^d \pi' \pi'' + \beta^e M(t, r), \nonumber\\	
\end{align} 
where 
\be
M(t,r) = \int_0^r 4 \pi r'^2 a(t)^2 \delta \rho(r',t) dr' \,,
\ee
is the enclosed mass within the radius $r$. Substituting these solutions into the scalar field equation, we obtain the equation solely written by the scalar field perturbations. The equation has the form given by
\be
(\gamma^a + \gamma^b M + \gamma^c M' )\pi' + \gamma^d \pi'^2 + \gamma^e \pi'^3 + \gamma^f M + \gamma^g \dot{M}=0.
\label{pi}
\ee
This is a non-linear algebraic equation for $\pi'$ and all the higher order derivative terms disappeared once the solutions for metric perturbations were substituted. 

We now introduce a mass dimension $\Lambda$ and assume the following scaling for the functions $G, A_3, A_4$ and $A_5$
\be
G \sim M_p^{2}, \;  X A_3 \sim X A_4 \sim X^2 A_5
\sim M_{p}  \Lambda^{-3},
\ee
where $M_p$ is the Planck mass and we assume $X \sim M_p \Lambda^3$. If the background scalar field is responsible for dark energy, then we expect $\Lambda^3 \sim H_0^2 M_P$ where $H_0$ is the present-day Hubble parameter.  
By introducing a new variable $x=\pi'/\Lambda^3 r$ and defining \cite{Kobayashi:2014ida}
\be
{\cal A} =\frac{M}{M_p \Lambda^3 r^3},
\ee
we obtain equation (\ref{pi}) in terms of the dimensionless field $x$, where $\pi'$ and $M$ are replaced by $x$ and ${\cal A}$ respectively.
We can now define the Vainshtein radius $r_V$ as the distance where, for $r < r_V$,  ${\cal A}$ becomes larger than unity, i.e.
$r{_V} =  (M/M_p \Lambda^3)^{1/3} $. For $r \ll r_V$ then, ${\cal A} \gg 1$ and $x \gg 1$. In this regime the solution for $\pi'$ is obtained as 
\begin{align}
\pi'^2 & = - \frac{\gamma^b M +\gamma^c M'}{\gamma^e}.
\end{align}
Substituting this solution into the metric perturbations, we obtain
\begin{align}
\Phi' & = \frac{G_N M}{r^2} + \frac{\Upsilon_1 G_N}{4} M'', \nonumber\\
\Psi' & = \frac{G_N M}{r^2} - \frac{5 \Upsilon_2 G_N}{4 r} M' 
+ \Upsilon_3 G_N M'',
\label{spherical}
\end{align}
where 
\begin{align}
\Upsilon_1 & = - \frac{(4 G_X - X A_3)^2}{4 A_3 G} , \nonumber\\
\Upsilon_2 &=   \frac{8 G_{X} X}{5 G},  \nonumber\\
\Upsilon_3 &=- \frac{- 16 G_X^2 + A_3^2 X^2}{16 A_3 G},
\label{Ypsilon}
\end{align}	
\begin{align}
G_N= \left[8\pi \left(2 G - 2 X G_{X} - 3 A_3 X^2/2\right)\right]^{-1},
\end{align}
and $X, G, G_{X}$ and $A_3$ are all evaluated at the background.
Outside a matter source ($M'=M''=0$), the solutions (\ref{spherical}) reduce to those in GR with a time dependent Newton constant $G_N$, thus the Vainshtein mechanism is working \footnote{These solutions need to be matched to the exterior solution at $r>r_V$ and this needs to be checked for a given choice of free functions. See \cite{Koyama:2015oma, Babichev:2016jom} for discussions.}. On the other hand, inside the matter source, the Vainshtein mechanism is broken and gravity is modified from GR. 

These results extend those obtained in \cite{Kobayashi:2014ida} for beyond Horndeski theory, which now (i.e. after the condition $A_1=0$ is imposed) corresponds to the following choice of the two free functions
\be
A_3 = - 4 G_{X}/X \,.
\label{HbH}
\ee
Imposing the above restriction, our results agree with those in \cite{Kobayashi:2014ida} where $\Upsilon_3$ vanishes. 

\noindent
{\bf Connection to effective theory of dark energy.}
On linear scales, cosmological perturbations are characterised by several functions of time within the framework of the effective theory of dark energy. In the Horndeski theory there are four parameters describing the nature of perturbations: $\alpha_M$, $\alpha_K$, $\alpha_B$ and $\alpha_T$ \cite{Bellini:2014fua}. These were extended to include beyond Horndeski theory with one more parameter $\alpha_H$ \cite{Gleyzes:2014rba}, and the degenerate higher order theories adding another parameter $\beta_1$ \cite{Langlois:2017mxy}. An interesting point is that the coefficients describing the spherically symmetric solutions (\ref{spherical}) can be written in terms of these parameters.  Expressing the latter in terms of $G$ and~$A_3$ we have
\be
\alpha_H= -\alpha_B= - 2 X \frac{G_{X}}{G}\,,  \quad
\beta_1 = \frac{X}{4 G} (4 G_X + X A_3 ) \,,
\label{alpha}
\ee
$\alpha_K$  and $\alpha_M$ do not contribute to the expressions (\ref{Ypsilon}), and clearly $\alpha_T=0$ by construction.
The violation of condition (\ref{HbH}) is indeed described by $\beta_1$. Note that the relation between $\alpha_B$ and $\alpha_H$ can be generalised by including the cubic Horndeski term. If $\beta_1 =0$, using (\ref{alpha}), it is easy to check that our results (\ref{spherical}) agree with those in \cite{Sakstein:2017xjx}.

\noindent
{\bf Observational constraints.}
In the case of $\Upsilon_3 =0$, interesting constraints on $\Upsilon_1$ and $\Upsilon_2$ have been obtained. $\Upsilon_1$ controls the modification of the Newton potential and the constraint comes from the structure of non-relativistic stars \cite{Koyama:2015oma, Saito:2015fza}. 
By demanding that the lightest observed red dwarf is at least as
heavy as the minimum mass for the onset of hydrogen burning in stars, a bound $\Upsilon_1 < 1.6$ was obtained in \cite{Sakstein:2015zoa, Sakstein:2015aac}. To constrain $\Upsilon_2$, we need relativistic observations. Comparing the weak lensing and X-ray mass of galaxy clusters, constraints $\Upsilon_1 = -0.11^{+0.93}_{-0.67}$  and $\Upsilon_2
= -0.22^{+1.22}_{-1.19}$ were obtained in \cite{Sakstein:2016ggl}. There is also a bound $\Upsilon_1 > -2/3$ coming from the fact that stable stars cannnot be formed if this bound is violated \cite{Saito:2015fza}. In the strong gravity regime, it was shown that the mass-radius relation of neutron stars is affected \cite{Babichev:2016jom}. Ref. \cite{Sakstein:2016oel} showed that the relation between the dimensionless momentum of inertia and the compactness of neutron stars is modified in beyond Horndeski theories and this relation is robust against the change of the equations of state. It will be interesting to revisit these studies in the presence of $\beta_1$.  

\noindent
{\bf Relation to Horndeski theory.}
Finally, we comment on the relation between the theory studied here, and the Horndeski one. It was shown in \cite{Crisostomi:2016czh, Achour:2016rkg} that theories in class N-I (which this theory belongs to) can be obtained from the generalised conformal and disformal transformation on the Horndeski theory  
\begin{equation}
\bar{g}_{\mu \nu} = \Omega(X) g_{\mu \nu} + \Gamma(X)
\phi_{\mu} \phi_{\nu}.
\end{equation}
The conformal transformation does not affect the propagation speed of gravitational waves while the disformal transformation does. Starting with a theory with $\bar{c}_{GW} \neq 1$, it is possible to perform a disformal transformation to make $c_{GW} =1$ by tuning $\Gamma$ and $\bar{G}$. The disformal transformation brings the Horndeski theory into beyond Horndeski theory \cite{Gleyzes:2014dya, Gleyzes:2014qga}. Thus, condition (\ref{HbH}) can be understood as this tuning. If matter was coupled to $\bar{g}$, the propagation speed of light was modified to $c =1/\bar{c}_{GW}$ and this did not change the ratio between the speed of gravity and light. However, we assumed that matter couples minimally to $g_{\mu \nu}$. Therefore, in the presence of matter, the theory we considered in this paper is different from the Horndeski one with minimally coupled matter, and this is the origin of the interesting phenomena concerning the breaking of the Vainshtein mechanism inside matter.  

\noindent
{\bf Discussions.}
In this paper, we studied the Vainshtein mechanism in the most general degenerate scalar-tensor theory propagating 3 dofs compatible with $c_{GW}=1$. This theory belongs to the class N-I \cite{Crisostomi:2016czh} (or Ia \cite{Achour:2016rkg}), with the additional condition that $A_1=0$ in the Lagrangian. The bottom line is that, excluding the cubic Horndeski case, the Vainshtein mechanism is irreversibly broken inside matter if the gravitational wave constraint is imposed. An interesting open question now concerns cosmology. It will be interesting to study the background expansion, as well as linear and non-linear structure formation in this theory, as the breaking of the Vainshtein mechanism can leave interesting imprints in large scale structure. 
 
\acknowledgments
\no MC is supported by the Labex P2IO and the Enhanced Eurotalents Fellowship. KK is supported by the STFC grant ST/N000668/1.  The work of KK and MC has received funding from the European Research Council (ERC) under the European Union's Horizon 2020 research and innovation programme (grant agreement 646702 ``CosTesGrav").

\end{document}